\documentclass[11pt]{article}
\usepackage{graphicx}
\usepackage{color}
\usepackage{bm}

\begin{document}
\title{Focusing Airborne Ultrasound with Partially Occluded Emission from Rectangular Plate with Flexural Vibration Mode}
\author{Keisuke Hasegawa*, Masaya Fujimori*, and Masaya Takasaki} 

\maketitle

\section*{Abstract}
We propose a focusing method of intense midair ultrasound out of ultrasonic emission from a single flexurally vibrating square plate partially covered with a purposely designed amplitude mask.
Many applications relying on nonlinear acoustic effects, such as radiation force employed in acoustic levitation, have been devised.
For those applications, focused intense airborne ultrasound is conventionally formed using phased arrays of transducers or sound sources with specific fabricated shapes.
However, the former strategies are considerably costly, and the latter may require minute three-dimensional fabrication processes, which both hinder their utility, especially for the construction of a large ultrasound emitting aperture.
Our method offers a possible solution for this, where the amplitude masks are designed in a fashion similar to the Freznel-zone-plate designing, but according to the positions of nodes and antinodes of the vibrating plate that are measured beforehand.

We experimentally demonstrate the successful formation of midair ultrasound focus at a desired position.
Our method only requires a monolithic plate, a driving transducer under the plate, and an amplitude mask fabricated out of laser machining processes of an acrylic plate.
Magnification of the spatial scale of ultrasound apertures based on our method is much more readily and affordably achieved than conventional methods, which will lead to new midair ultrasound applications with the whole room workspace.
\footnote{K. H and M. F equally contributed to this paper.\\K. H. and M. T, Authors are and with Department of Mechanical Science, Graduate School of Science and Engineering, Saitama University, 338-8570, Saitama, Japan.\\
M. F. was with the Department of Mechanical Engineering and System Design, Faculty of Engineering, Saitama University.
K. H. Author is to whom corresponding should be addressed (e-mail: keisuke@mech.saitama-u.ac.jp).\\
This project is financially supported by Japan Society for the Promotion of Science (JSPS), Grant No. 23K18488, Japan.}

\section{Introduction}
\label{sec:introduction}
\subsection{Prevalent use of focused midair ultrasound and current focusing methods}
Applications of midair focusing ultrasound have been continuously investigated, especially regarding the ones that rely on nonlinear acoustic effects exhibited by intense ultrasound fields, such as acoustic radiation force \cite{af1, af2}, acoustic streaming \cite{as1, as2, as3}, and parametric array \cite{ps1, ps2}.
The unique nature of focused intense midair ultrasound that it can engender non-contact mechanical effects at an arbitrary pinpoint position in the air has materialized many practical applications such as acoustic levitation \cite{al1, al2, al3}, midair object manipulation \cite{om1, om2}, non-contact tactile displays \cite{td1, td2, td3}, midair fragrance transportation \cite{frag}, remote suppression of short-range exposure to infectious aerosols \cite{aerosol}, and so forth.
On the other hand, achieving sufficiently strong focusing of midair ultrasound requires engineering of the emitted wavefront, which is not possible with a single transducer alone, in general.
To this end, the following two strategies have been adopted.

One is the use of phased arrays of airborne ultrasound transducers.
Since its advent \cite{tatezono, hoshi}, many applications have been devised based on this technique.
The most prominent advantage of this approach is that ultrasound fields with a huge variety in spatial distributions can be promptly generated by electronically controlling the phase delays of emissions from individual transducers.
This feature is particularly beneficial for applications that require real-time reconfiguration of the generated ultrasound field.
However, this approach requires numerous transducers to form a satisfactorily large ultrasound emission aperture.
Most current airborne ultrasonic phased arrays employ cylindrical transducers with a resonance frequency of 40~kHz because of their relatively low attenuation through propagation.
Among commercially available products, the most common transducers of that resonance frequency are those with a diameter of 10~mm \cite{aupa1, aupa2}.
When constructing an emission plane using those transducers, 100 pieces of them are required for a 100~mm $\times$ 100~mm aperture.
Implementing arrays of such a large number of transducers with their emission individually controlled requires much work and cost, especially when one wants a large aperture that can focus over a long distance from the emission plane.

The other approach is physically forming an emission plane with a spatial distribution of output phase or amplitude.
The biggest disadvantage of this approach is that the focal positions cannot be electronically changed as with the phased array technique.
Nevertheless, this approach is generally more affordable and is easier to fabricate than the previous one.
The strategies for realizing planar distributions of output phases include emission surface with designed uneven height \cite{uneven1, uneven2} and use of acoustic metasurfaces or metamaterials \cite{am1, am2, am3, am4}.
Although those methods do not require arrays of emission-controlling circuits, they require relatively complicated and minute fabrication processes, sometimes in a three-dimensional fashion.

\begin{figure*}[!t]
\centerline{\includegraphics[width=\textwidth]{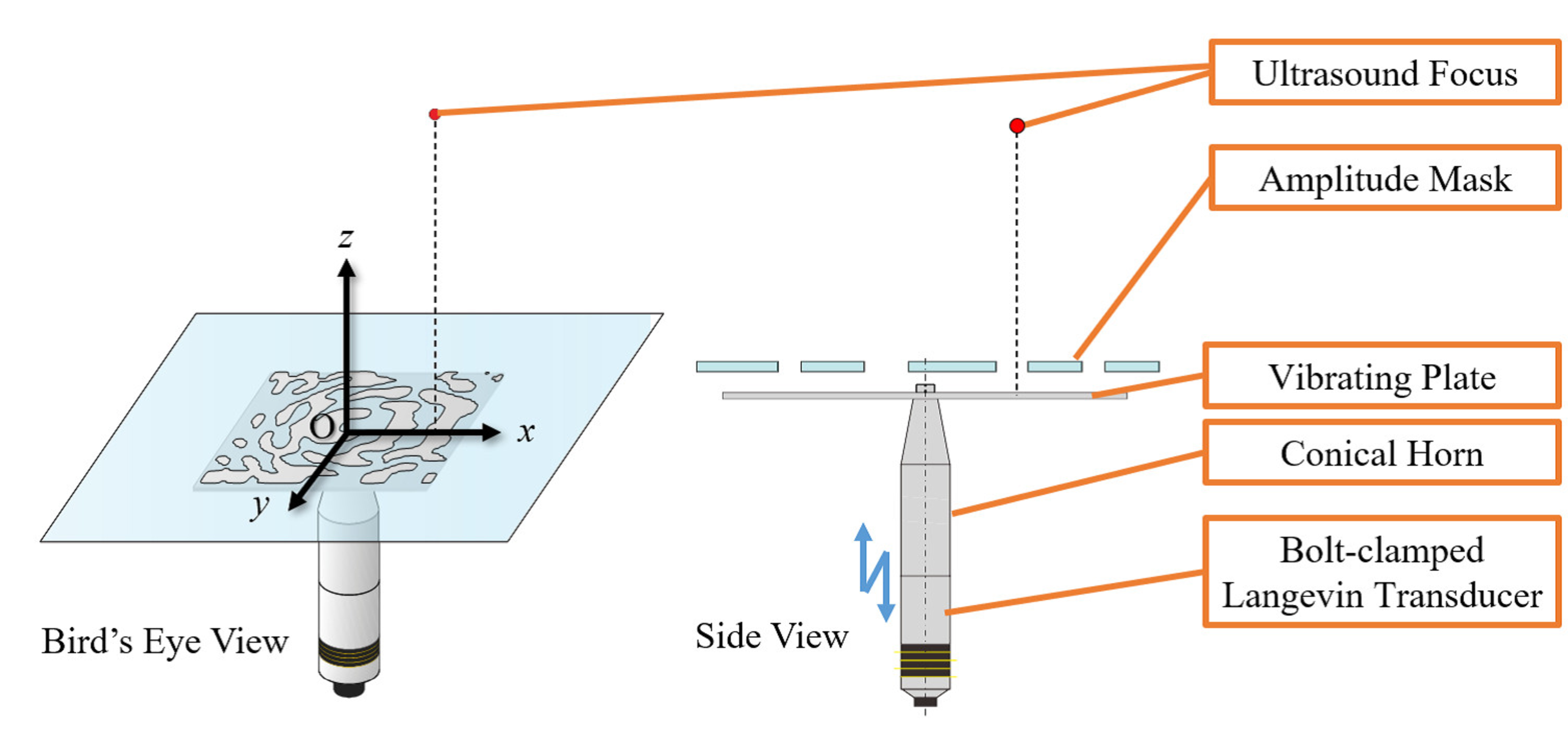}}
\caption{Schematic depiction of our proposed strategy for creating a midair ultrasound focus out of partially occluded ultrasound emission from a flexurally vibrating plate.}
\label{fig:first}
\end{figure*}

\subsection{Focusing via Amplitude Distribution of Emission and Remaining Problems}
 Contrastingly, creating amplitude distribution on the emission plane is relatively easy, where a planar incident wave is transformed into a desired wave field via an amplitude mask with a spatial distribution of acoustic transmittance.
A concentric transmittance distribution known as the Freznel-zone-plate (FZP) is a representative approach, which offers focusing of planar incident waves on the center axis of the plate with a desired focusing depth \cite{fzp1, fzp2, fzp3, fzp4}.
However, this approach assumes that the incident wave is planar.
Therefore, a considerable space between the wave source and the FZP-based amplitude masks is required, which results in the need for a considerable space behind the mask in creating a large ultrasound emission aperture based on this approach.

To overcome this problem, we previously demonstrated an FZP-based focusing method, where an array of airborne ultrasound transducers all driven in the same phase serve as a spacious source of a planar wave \cite{kitano}.
An FZP is placed just above the transducers to generate an intense midair ultrasound focus in the same way as with methods employing FZPs for planar incident waves.
This method can construct focusing ultrasound apertures with side lengths of several hundreds of millimeters with a configuration where virtually no space between the ultrasound source and the amplitude mask is required.
A planar array of transducers driven in phase was adopted to realize an approximated planar wave incidence with that large aperture, just above the emission plane.
The FZP here is fabricated from an acrylic plate that was processed with a laser-machining apparatus with millimeter precision, which is much more affordable than the processing and materials required for fabricating planes with uneven height or acoustic metamaterials.
However, arranging thousands of transducers is still costly and may be an obstacle that may prevent this technique from being applied to the fabrication of an even larger emission plane.
If such a huge emission plane for an intense midair ultrasound field is realized, constructions of upsized versions of all current application systems to a whole-room scale can be materialized, by which many new application scenarios will emerge.

\subsection{Motivation and Contribution of the Study}
 The goal of this study is to establish a method that can potentially create a meter-order ultrasonic emission plane generating a midair ultrasound focus at a desired position.
To this end, we conceived of utilizing a monolithic vibrating plate as an ultrasound source and partially covering its emission similarly to the FZP-based techniques (Fig. \ref{fig:first}).
Creating such a vibrating plate is so readily done with a single transducer placed on the plate, and the plate can be affordably upsized.
In this study, we drive a square duralumin plate supported by a single Langevin ultrasonic transducer driven in longitudinal vibration mode.
This construction strategy of ultrasound emission from a rigid body where a single Langevin transducer is attached to a plate is found in preceding studies \cite{plate, plate2}.
The novelty of this study is that a newly defined problem is handled where a midair ultrasound focus is to be created out of incident waves that have phase and amplitude distribution on the amplitude mask, which correspond to the positions of nodes and antinodes on the vibrating plate.

 The technical contribution of this paper is as follows:
 \begin{itemize}
     \item conceptualization of the aforementioned new focusing method based on partial occlusion of emission from a flexurally vibrating plate
     \item derivation of a designing method of amplitude masks over the plate
     \item experimental validation of successful midair focusing based on the proposed strategy
 \end{itemize}
In the following parts of the paper, we first describe the physical modeling of the flexurally vibrating plate as an ultrasound emission source.
Next, we derive a design algorithm for amplitude masks based on this modeling.
Then, we explain the experimental apparatus composed of a vibrating plate activated by a Langevin ultrasound transducer supporting the plate from its bottom and fabricated amplitude mask covering the plate.
Finally, we show the result of ultrasound field measurement, in which midair ultrasound foci are successfully generated as intended at desired positions.
We expect that the proposed method can potentially be one of the new standard methods for system construction for generating intense midair focused ultrasound because of its simplicity, affordability, and readiness for being upsized.
 
\section{Method}
\subsection{Modeling of Ultrasound Emission Sources}
 We begin with modeling the flexural vibration plate as a two-dimensional array of point sources of pressure vibration.
A plate vibrating with a single certain flexural mode contains nodes and antinodes on its surface.
We assume that the individual equivalent point sources on a plate have their emission amplitudes being proportional to the vibrating amplitude of the plate at the source positions.
We also assume that the emission phases of the sources are the same as those of the vibration of the plate at the source positions.
The validity of this assumption is later demonstrated by acoustic measurement and numerical simulation results.
 \subsection{Designing Method of Amplitude Masks for Focusing}
 Let $\bm{r}_t = (x_t, y_t, 0)$ be a position of the equivalent point source on the plate and $\theta_t(\bm{r}_t)$ be the emission phase delay of the source at $\bm{r}_t$, where the coordinate system is defined so that the origin is located on the center of the plate surface (Fig. \ref{fig:first}).
 Here, the value of $\theta_t(\bm{r}_t)$ is ideally either 0 or $\pi$ (note that if this is not the case for actual vibrators, the following method for designing amplitude masks is still valid.)
 When the desired focal position is given as $\bm{r}_f = (x_f, y_f, z_f)$, the criteria of creating an amplitude mask is realizing a large ultrasound amplitude at $\bm{r}_f$ out of partial emission from the sources in a manner that those emissions can constructively interfere at the focal position.

 With $k$ denoting the wavenumber of ultrasound emission in the air, the phase delay of ultrasound observed at the focal position, which is emitted from a source located at $\bm{r}_t$ is given as
 \begin{equation}
 \theta_d(\bm{r}_t) = k||\bm{r}_t - \bm{r}_f||,
 \end{equation}
 where $||\cdot||$ denotes the vector norm of $\cdot$.
 In conventional FZP-based mask designing procedures, the value of $\theta_t(\bm{r}_t)$ is uniform at any position on the plate.
 For this case, the amplitude mask is designed with a simple criterion: according to quantization of $\theta_d(\bm{r}_t)$ by $\pi$ after phase wrapping, where to perforate in the mask is determined.
 This situation is understood as that the emissions with similar phase delays at the focal points are permitted, resulting in intense ultrasound amplitude exclusively at the focal position.
 Concentric mask patterns commonly created on FZPs are the outcome of those phase-quantization procedures.
\begin{figure*}[!t]
\centerline{\includegraphics[width=\textwidth]{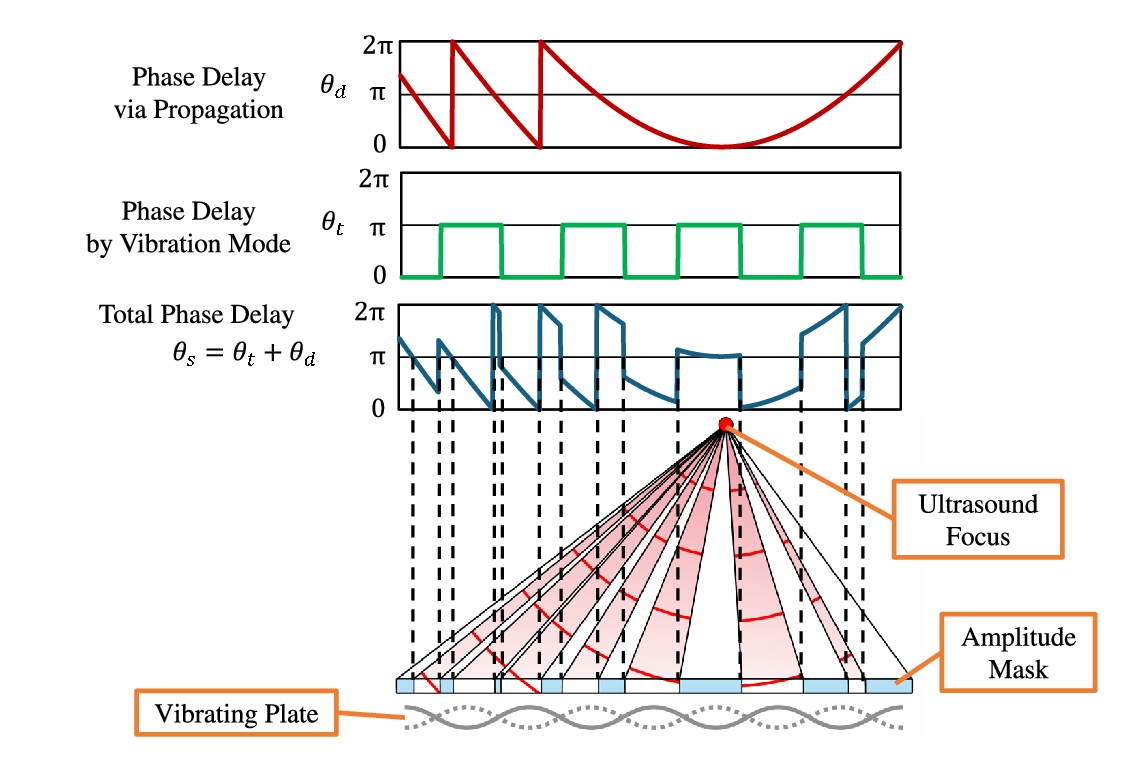}}
\caption{The designing method for the amplitude mask based on the calculation of phase delays at the focal position from each emission point. All phase graphs are drawn after being wrapped by $2\pi$.}
\label{fig:calc}
\end{figure*}
 Due to the flexural vibration on the plate yielding position-dependent emission phase delays, in this study, the phase delay of the emissions from $\bm{r}_t$ at the focus is not equal to $\theta_d(\bm{r}_t)$.
 The total phase delay $\theta_s(\bm{r}_t)$ observed at the focal position, is given as the sum of the emission phase delay $\theta_t(\bm{r}_t)$ and propagation phase delay $\theta_d(\bm{r}_t)$. Therefore,
\begin{equation}
\theta_s(\bm{r}_t) = \theta_t(\bm{r}_t) + \theta_d(\bm{r}_t)
\end{equation}
 holds.
 The above quantization criterion is again adopted here, finally giving a design rule for binary mask amplitude distributions $P(\bm{r}_t)$:
 \begin{equation}
 P(\bm{r}_t)  =  \left\{
 \begin{array}{cc}
 1, & 2n\pi \leq \theta_s(\bm{r}_t) - \alpha < (2n + 1)\pi\\
 0, & (2n + 1)\pi \leq \theta_s(\bm{r}_t) - \alpha < 2(n + 1)\pi
 \end{array}
 \right. ,
 \end{equation}
 where $\alpha \in [0, 2\pi]$ denotes a threshold value of the phase and $n$ denotes an integer that can be any of negative, zero, or positive.
A designed amplitude mask is supposed to block off the incident ultrasound at $\bm{r}_t$ where $P(\bm{r}_t) = 0$, and to let the ultrasound travel through it where $P(\bm{r}_t) = 1$.
Note that the values of $\theta_d(\bm{r}_t)$ and $\theta_s(\bm{r}_t)$ here are not wrapped by $2\pi$.
Here, the absolute (not relative) value of $\theta_s(\bm{r}_t)$ is essentially indeterminable: an arbitrary constant phase shift can be added on all values of $\theta_t(\bm{r}_t)$ because there is no phase origin for $\theta_t(\bm{r}_t)$.
Also, the resulting focal intensity relies on values of $\alpha$.
Therefore, in this study, we defined the value of $\theta_t(\bm{r}_t)$ as phase delays of the surface vibration from the driving signal of the transducer, which is measured in an experiment described in the following section.
We set the value of $\alpha$ so that the focal amplitude becomes the largest, which is calculated with the numerical simulation method described in section 4 while varying the value of $\alpha$.
The whole calculation processes are schematically depicted in Fig. \ref{fig:calc}.

 \section{Experimental Setting}
\subsection{Composition of Experimental Apparatus}
\begin{figure}[!t]
\centerline{\includegraphics[width=0.6\textwidth]{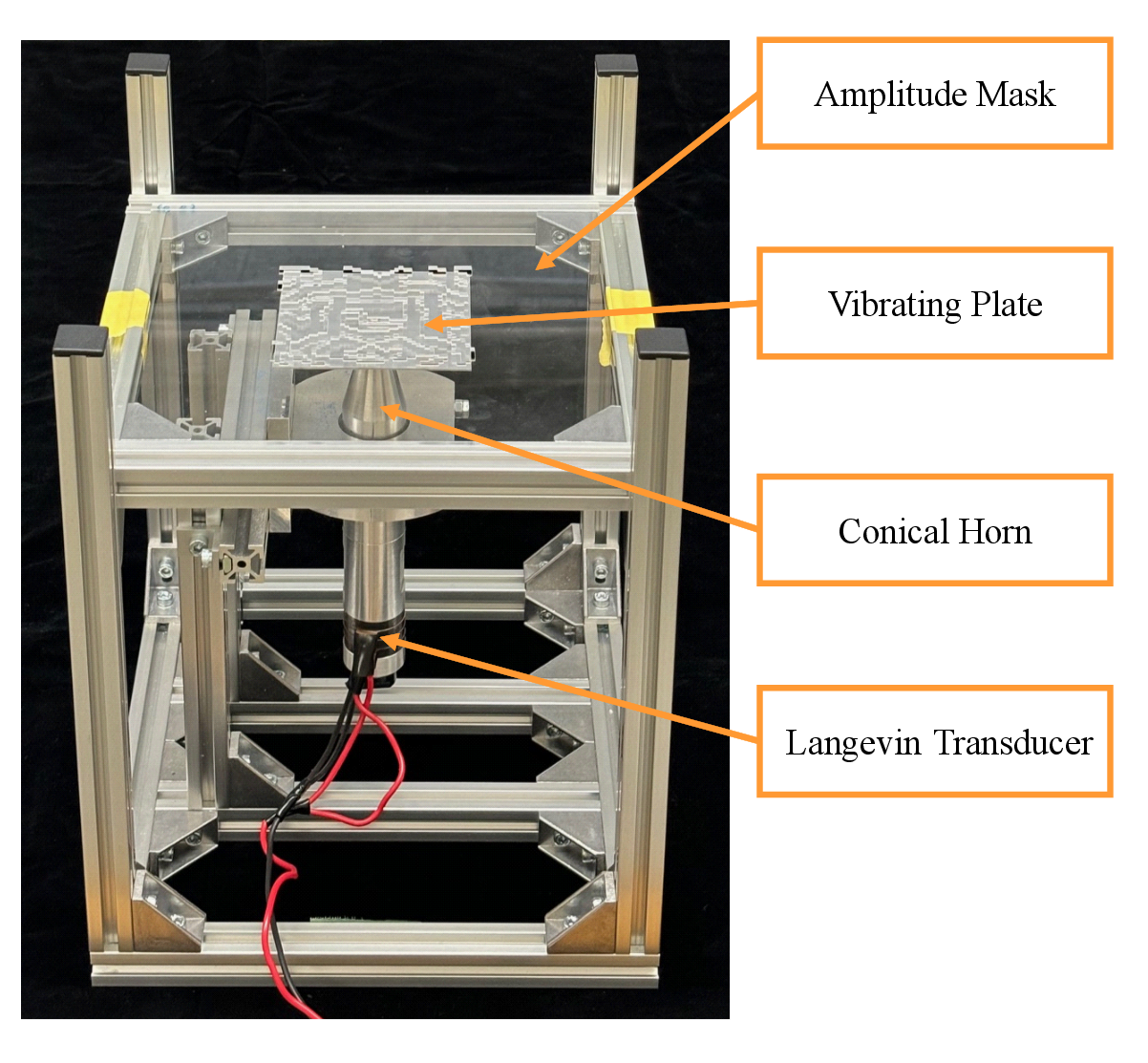}}
\caption{Fabricated experimental apparatus.}
\label{fig:apparatus}
\end{figure}

Figure \ref{fig:apparatus} shows the constructed experimental apparatus containing a cylindrical Langevin ultrasonic transducer with a resonance frequency of 28~kHz (HEC-3028P4B, Honda Electronics Co., Ltd., Japan), a square duralumin vibration plate with a thickness of 1~mm, and a conical acoustic horn made of duralumin that combines the tip of the transducer and the vibrating plate as ultrasound emitting components.
The top and bottom diameters of the horn were 6~mm and 30~mm, respectively.
The center of the vibration plate contained a through hole of 3.2~mm in diameter, where the horn tip was bolted. 
Those components were mounted on an external aluminum frame via a metallic jig that was in contact with the sides of the horn on the nodal position to supress vibration absorption by the frame.
A fabricated amplitude mask was placed on the top of the frame with no direct contact with the vibrating plate.
\subsection{Dimension Design of the Vibrating Components}
According to the resonance frequency of the transducer, we determined the dimensions of the acoustic horn and the vibrating plate using numerical simulations based on the finite-element method.
The length of the horn was determined so that a single node where the holding jig should be in contact was exhibited at the vibrating mode of 28~kHz.
The side length of the square vibrating plate was determined as 87.6~mm.
We experimentally verified that the whole vibrating system where the transducer, the acoustic horn, and the vibrating plate were coalesced, had its forced resonance output peak around the resonance frequency of the transducer.
\subsection{Fabrication of the Amplitude Masks}
We adopted a 2-mm-thick acrylic plate as the material for the amplitude mask.
We cut out the plate using a laser machining apparatus (Etcher Laser Pro, smartDIYs Co., Ltd., Japan).
The mask pattern was constructed by an array of square perforations with a side length of 2~mm, which is calculated from the result of the following source phase distribution measurement and phase quantization formulated in eq. (3).
Adjacent perforations are incorporated to form a large transmission area in the plate.
In addition, several parts that were isolated from the edge of the pattern as a result of pattern calculation were manually adjusted so that they were in contact with the plate.

\subsection{Measurement Procedures}
\begin{figure}[!t]
\centerline{\includegraphics[width=\textwidth]{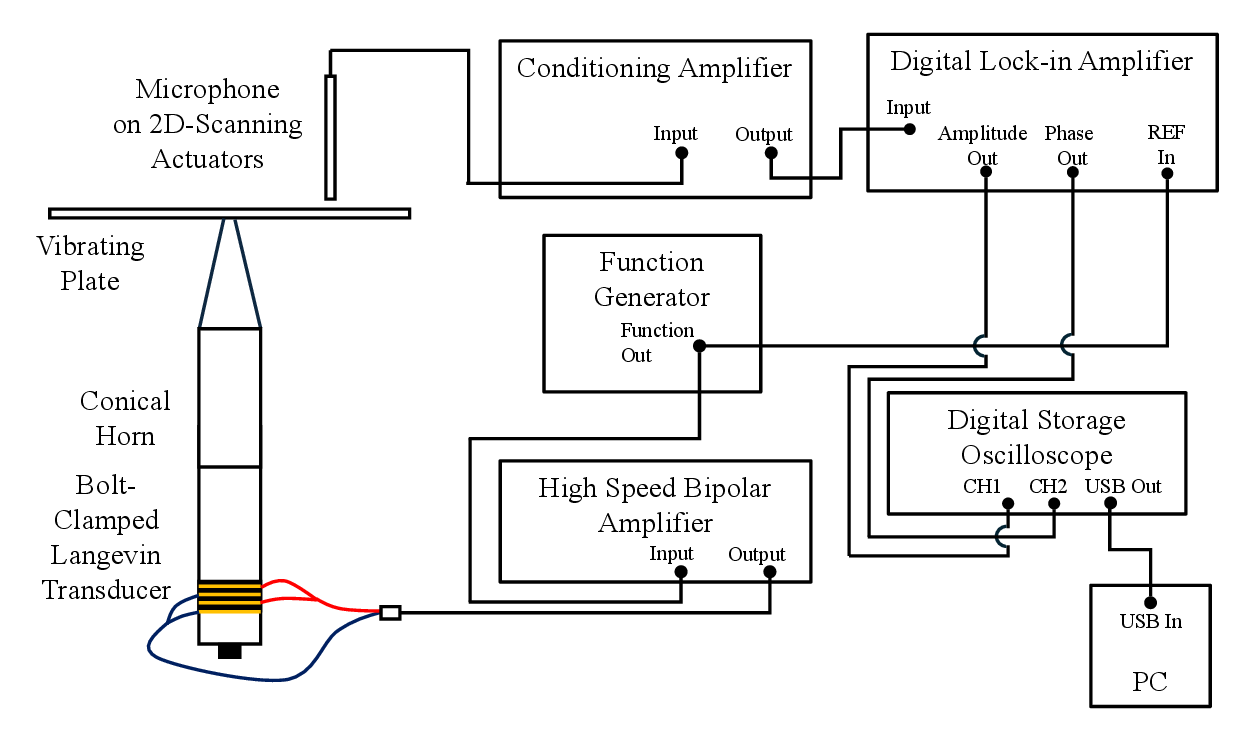}}
\caption{Schematic depiction of the experimental setup and signal flows in acoustic measurement.}
\label{fig:setup}
\end{figure}

We set the driving frequency of the vibration system to 27,877~Hz throughout the following acoustic measurement experiments.
We used a standard microphone system (1/8-in. microphone, type 4138-A-015; pre-amplifier, type 2670; condition amplifier, type 2690-A; all products of Hottinger, Br\"{u}el and Kj\ae r, Denmark).
This microphone has an almost flat frequency response from 6.5~Hz to 140~kHz.
The directivity of the microphone is limited within 3~dB across the elevation angle of the incident wave ranging from 0 to 90 degrees.
The measured signal was sent to a lock-in amplifier (LI5640, NF Corporation, Japan) to calculate the phase and amplitude of the frequency component in the measurement that matched the sinusoidal driving signal provided by a function generator as a reference signal.
The driving signal was also provided from the same function generator in parallel, and was amplified via a bipolar amplifier (HSA4052, NF Corporation, Japan), and the driving voltage of the transducers was set to 3.6~V in RMS value.
The microphone scanned two-dimensionally on a plane that was parallel to the xy-plane defined by the coordinate system in Fig. \ref{fig:first} using combined two-axis linear actuators (EZSM4D030AZAC
 and EZSM6D070AZAC, Oriental Motor Co., LTD., Japan).
 The whole experimental setup is schematically depicted in Fig. \ref{fig:setup}.
 We used a PC for the storage of data that was transmitted from a digital storage oscilloscope.

\section{Experimental Results}

\begin{figure}
\centerline{\includegraphics[width=0.45\textwidth]{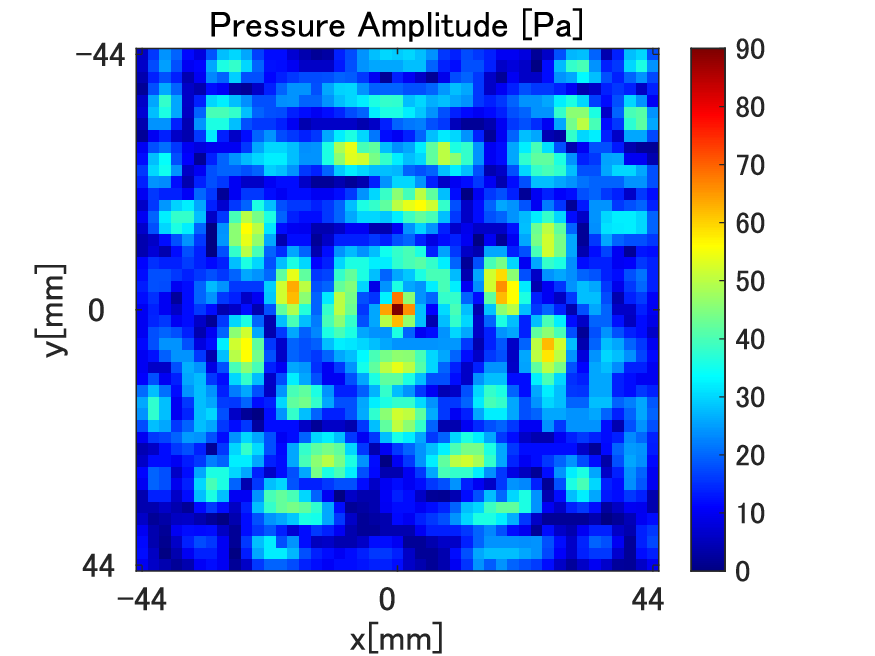} \includegraphics[width=0.45\textwidth]{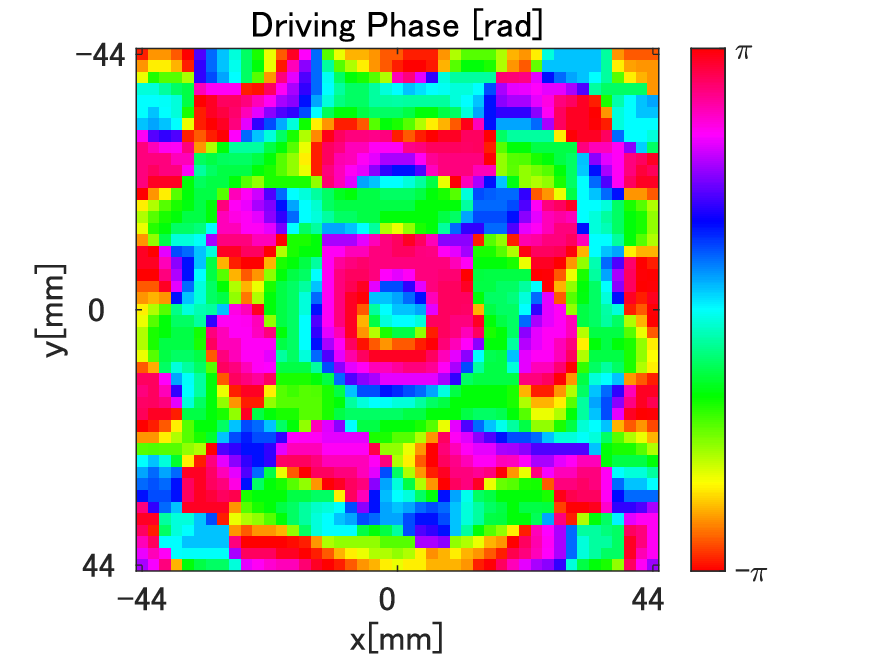}}
\caption{Measured pressure amplitude distribution (Upper Figure) and phase distribution (Lower Figure) near the vibration surface of the plate.}
\label{fig:surface}
\end{figure}

\subsection{Surface Vibration Measurement of the Plate and Fabrication of the Amplitude Masks}
We first measured surface pressure distribution just above the vibrating plate, which is necessary for the designing of the amplitude mask.
We scanned a plane located 2~mm above the plate with spatial intervals of 2~mm, corresponding to the spatial resolution of the mask-designing procedure.
The size of the scanning area was 88~mm on each side.

\begin{figure}
\centerline{\includegraphics[width=\textwidth]{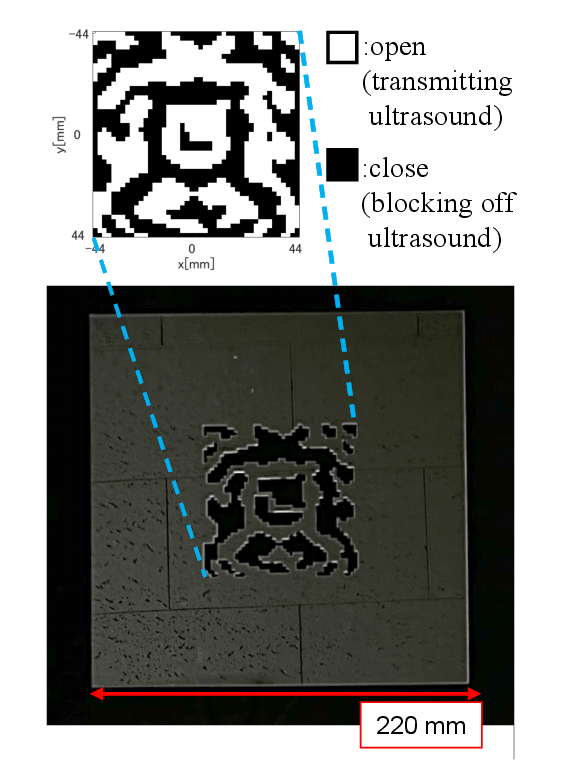}}
\caption{Example of mask pattern calculation (upper figure) and fabricated acrylic amplitude mask plate (lower figure) for the focal position of $\bm{r}_{f1}$ = (0~mm, 0~mm, 100~mm). White and black regions in the pattern correspond to the region where ultrasound is transmitted and is blocked off, respectively. As stated in the main text, bridging 'close' elements that were not indicated in the mask patterns in the upper figure were manually added to actual masks to prevent some 'close' regions from being isolated from the surrounding parts of the plate. }
\label{fig:maskexample}
\end{figure}

\begin{figure*}[!t]
\centerline{\includegraphics[width=\textwidth]{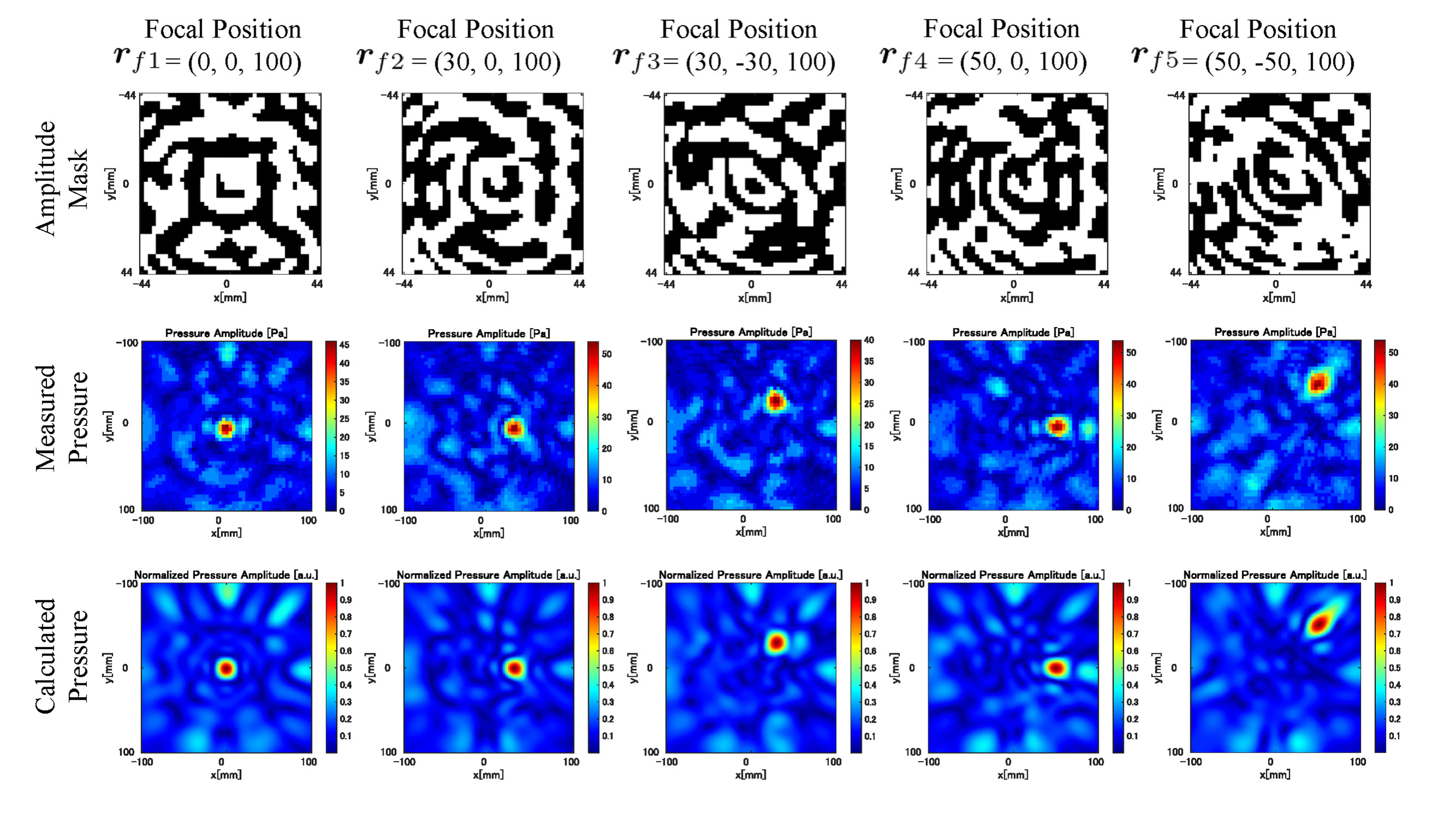}}
\caption{Calculated amplitude mask patterns (upper row), measured ultrasonic pressure amplitude distributions (middle row), and numerically calculated pressure amplitude distributions (lower row). Each column corresponds to the focal possition of $\bm{r}_{f1},\ldots, \bm{r}_{f5}$ from left to right. All focal coordinates are in millimeters.}
\label{fig:meas}
\end{figure*}

Figure \ref{fig:surface} shows the obtained amplitude and phase distributions of pressure waveform over the vibrating plate.
Note that the phase distribution shown here is not the phase delay of the plate vibration but the phase advance of that compared with the reference signal fed to the lock-in amplifier.
Based on this measurement, we created five amplitude masks as shown in Fig. \ref{fig:maskexample}.
The focal positions are $\bm{r}_{f1} = (0,0,100),\bm{r}_{f2} = (30,0,100),\bm{r}_{f3} = (30,-30,100),\bm{r}_{f4} = (50,0,100),$ and $\bm{r}_{f5} = (50,-50,100),$ all indicated in millimeters.
In the upper row of Fig. \ref{fig:meas}, we indicate all calculated amplitude mask patterns for the focal positions from $\bm{r}_{f1}$ to $\bm{r}_{f5}$.

\subsection{Acoustic Field Measurement and Numerical Simulations}

Next, we placed each amplitude mask over the vibrating plate and scanned the microphone in a square region with a side length of 200~mm, which was located 100~mm above the plate with a scanning interval of 4~mm.
Because the wavelength of 27,877~Hz ultrasound in air is 12.2~mm, we consider this scanning interval to be sufficiently small.
The size of the scanning area was 200~mm on each side.

The measurement results for each amplitude mask are shown in the middle row of Fig. \ref{fig:meas}.
Note that the captured waveforms included weak harmonics of the driving frequency of the transducer due to the nonlinearity of air.
Nevertheless, the RMS values of the captured waveforms and that of the driving frequency component in that waveform extracted by the lock-in amplifier showed little difference.
Therefore, we show the latter values as pressure amplitude in the figures.
At the intended focal position, intensified ultrasound amplitude is observed for all cases.
The diameters of the foci are approximately 20-30~mm.
Around the foci, several artifacts are observed whose intensities are not negligibly small.

To verify the assumption that the flexural vibrating plane can be modeled as a set of small point sources with phase distributions, we performed numerical simulations of generated ultrasound field.
By assuming omnidirectivity of the point sources, based on the principle of superposition, the resulting complex amplitude field $p_{\mathrm{sim}}(\bm{r})$ with respect to $\bm{r}$ denoting an arbitrary position in the air is given as
\begin{equation}
p_{\mathrm{sim}}(\bm{r}) = \sum_m A_m\frac{e^{-jk||\bm{r} - \bm{r}_m||}}{||\bm{r} - \bm{r}_m||},
\end{equation}
where $j = \sqrt{-1}$ is the imaginary unit, $m$ is an index of point sound sources, $A_m$ is complex emission amplitude of the source, and $\bm{r}_m$ is the source position.
We calculated the amplitude distributions for each amplitude mask based on Eq. (4), with the values of $A_m$ determined according to the surface pressure measurement shown in Fig. \ref{fig:surface} and the masking pattern.

The lower row of Fig. \ref{fig:meas} shows the simulation results.
It is confirmed that the results of simulations and measurements show good agreement.
These results indicate that the point source modeling approach adopted in our study faithfully reflected actual physical phenomena engendered with our apparatus.
It is also demonstrated that the aforementioned simulation procedures can accurately predict the resulting ultrasound field out of a pair of a vibration mode of a plate and an amplitude masking pattern on the plate.

\section{Discussions}
\subsection{Artifacts Superimposed onto the Desired Focal Amplitude Distribution}
\begin{figure}
\centerline{
\includegraphics[width=0.5\textwidth]{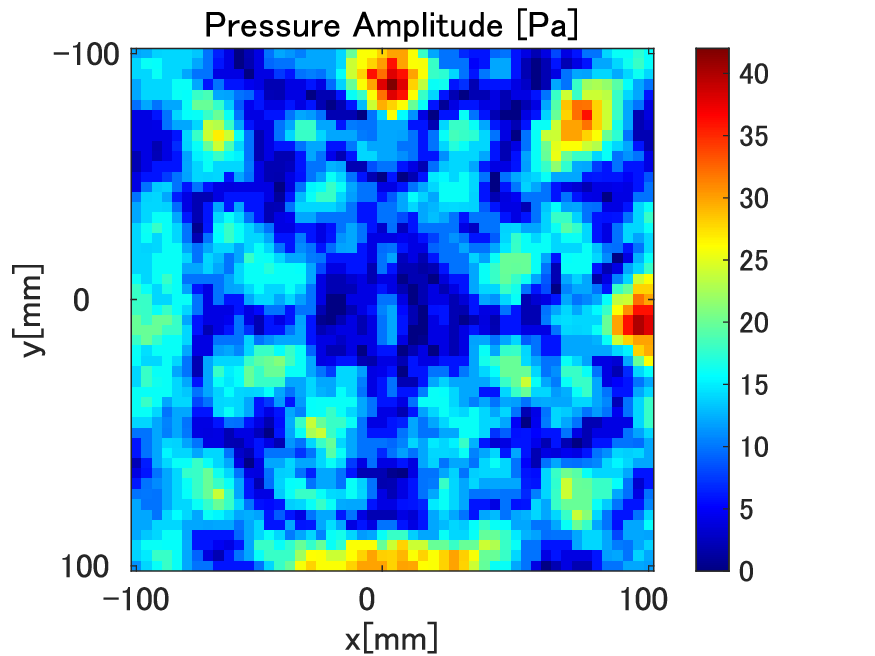}
\includegraphics[width=0.5\textwidth]{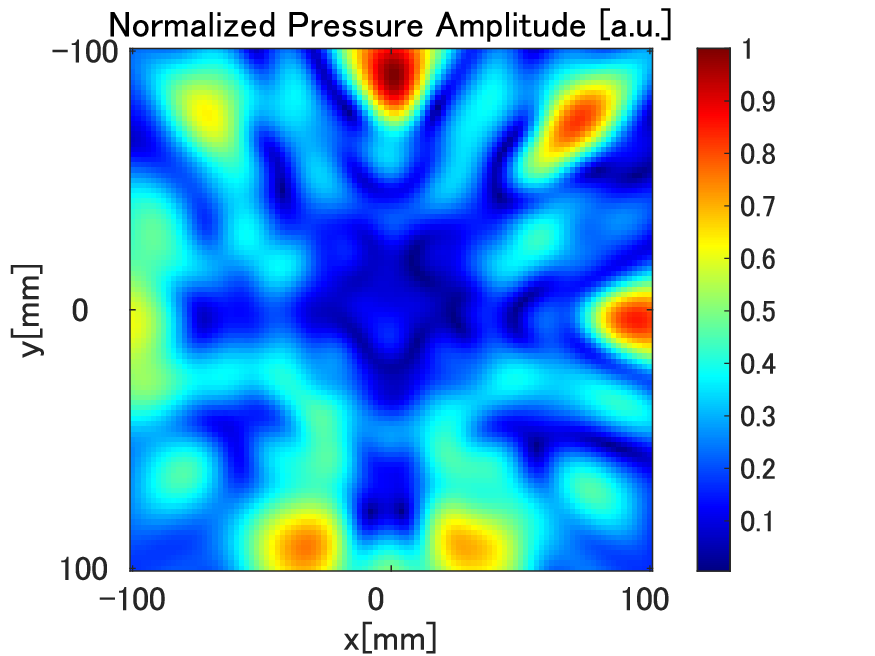}
}
\caption{Pressure amplitude pattern obtained via measurement (left figure) and numerical simulation (right figure) where no amplitude mask is placed over the vibrating plate.}
\label{fig:nofzp}
\end{figure}

As stated, the intensity of some artifacts is relatively large, which may be undesirable in practical situations.
Figure \ref{fig:nofzp} shows the measurement and numerical simulation results of ultrasound amplitude distribution where no amplitude mask was placed on the vibrating plate.
The results show that the artifact patterns around the foci are similar to this mask-free amplitude pattern.
This fact is understood as follows.
partially covering the emission from the plate is equivalent to adding some spatial interference to the original emission pattern of the plate shown in Fig. \ref{fig:nofzp}.
This spatial interference can create a new focusing point as desired, but will not completely delete the original amplitude image, which inevitably is superimposed to the focal pattern.

\begin{figure}
\centerline{\includegraphics[width=0.5\textwidth]{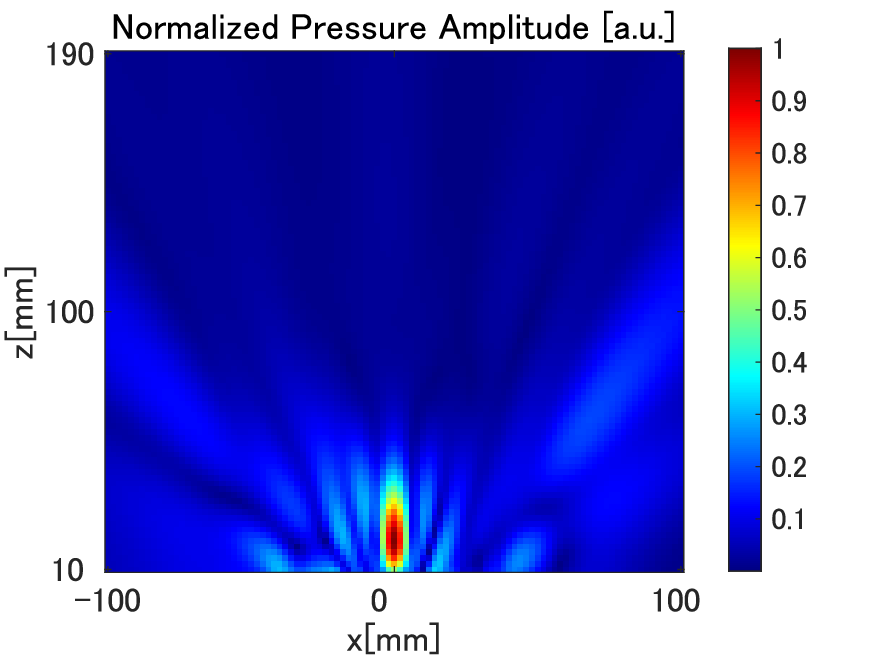}
\includegraphics[width=0.5\textwidth]{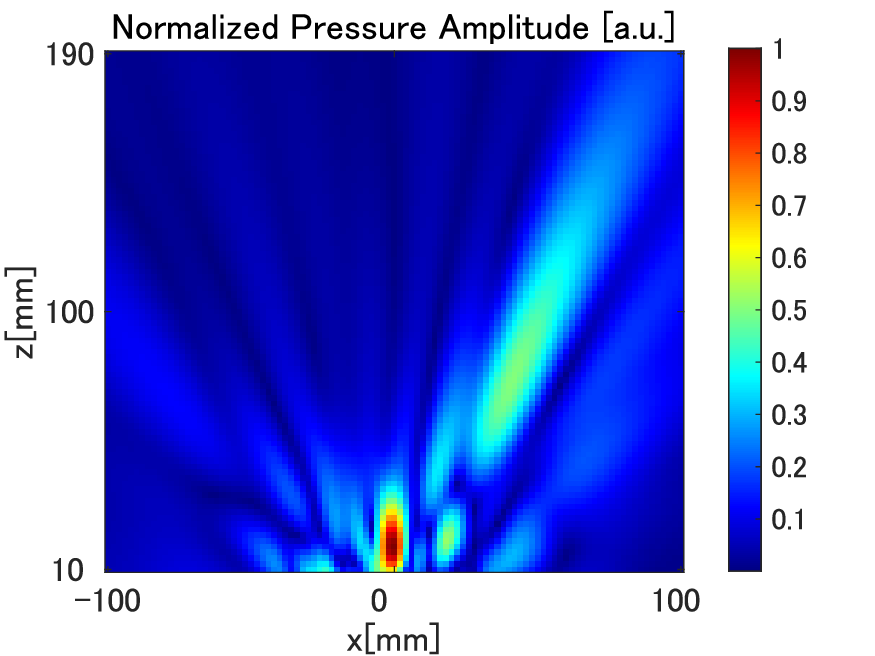}}
\caption{Pressure amplitude simulation on the $xz$-plane with no amplitude mask placed over the vibrating plate (left figure), and that with the amplitude mask for the focus at $\bm{r}_{f4}$ (right figure).}
\label{fig:side}
\end{figure}

Next, we show the numerical simulation results for amplitude distribution on the $xz$-plane in Fig. \ref{fig:side}, which was unable to be scanned with the actuators that we had.
According to the results, a small region with intense ultrasonic power conversion is observed just above the center of the plate.
This is because the vibration mode had a spatial phase distribution similar to concentric rings, which was expected to yield an on-axis constructive ultrasound beam like the Bessel beam \cite{as3}.

The resulting vibrating mode on the plate may be somewhat inappropriate in a practical context because of a few artifacts.
A different vibrating mode such as the one with checker-board-like phase distribution may exhibit much weaker artifacts around the focal position.
In addition, it may be the case that the artifact could be partially explained by the grating lobe effect of arrayed emitters where unintentional radiation directivities occur in certain directions due to spatial intervals between array elements longer than half the wavelength.
Therefore, a two-dimensional phase distribution mode with sufficiently short inter-node distances would be one good solution for reducing artifacts and thereby improving the focusing efficiency.
\subsection{Limitation and Possible Improvement of Current Apparatus for Practical Use}
The current apparatus drives the vibrating plate via a very small single-bolted joint surface attached to the acoustic horn.
We found that excitation via this joint surface with higher driving voltage often resulted in the gradual relaxation of the bolt.
Due to this, the whole experiments were conducted while suppressing the excitation voltage within relatively a small value.

Because the main motivation of our experiments is to experimentally demonstrate the validity of our strategy of partial covering of flexural emission for midair ultrasound focusing, the emission aperture in the experiments was kept relatively small.
If one wants to form a larger emission plane out of our technique, the single-point support of the vibrating plate adopted in the experiment is not suitable because the plate will become heavier for larger apertures.
To achieve a more stable operation with such a large vibrating plate, driving multiple antinodes of the desired mode with several harmonically driven transducers that also support the vibration plate may be a solution for this problem.

Another important aspect is the choice of the plate material.
Current material, the aluminum plate might not be the best choice when the plate becomes larger and is driven with greater displacement.
For those cases, thermal dissipation inside the plate may be crucial, so more efficient (lossless) material may be needed to be used.
As another solution for this, attaching small thermal components serving as heat sinks on nodes of the plate could be applicable for promoting heat convection to the air without lowering the vibration power of the plate.


\section{Conclusions}
In this study, we proposed a method that enables the construction of an ultrasound emission aperture for creating a midair ultrasonic focus that is composed of a monolithic vibration plate and a planar amplitude mask that partially occludes the emission of the plate.
In designing processes of amplitude masks, we modeled the emission from the plate as a set of small point sources and derived an amplitude distribution calculation method based on phase-delay quantization that is an extension of the conventional FZP designing criteria.
We fabricated amplitude masks based on that procedure and placed them above the vibrating plate to create midair foci of ultrasound.
We measured the generated ultrasound amplitude distribution and demonstrated the successful generation of ultrasound foci at intended positions.
Results of numerical simulations and acoustic measurement showed good agreement, which indicates the validity of our modeling strategy of a vibrating plate as a set of point sources and the designing algorithm of amplitude masks.

The aperture of the current experimental apparatus is 87.6~mm in side length, and the focal depth is only 100~mm.
To fabricate an upsized version, supporting and activating the plate with multiple synchronously driven transducers is one of the reasonable solutions.
Even in those cases, the proposed method is expected to be much more affordable than conventional methods, where minute fabrication of acoustic metamaterials or compactly arranged thousands of transducers in the emission plane are required.
We consider that the focusing strategy demonstrated in this paper is currently one of the most realistic potential compositions of meter-order emission apertures.
It is expected that new applications such as acoustic levitation or noncontact actuation in a whole room scale can be materialized based on our technique in the future.
To demonstrate the possibility of such applications, the physical construction of a larger emission plane remains our next challenge.
From a viewpoint of practicality, establishing a method of electronically controlling amplitude distribution on the mask is also an important issue.


\begin{thebibliography}{00}

\bibitem{af1} L. V. King, "On the Acoustic Radiation Pressure on Spheres, " Proc. R. Soc. Lond., A 147, pp. 212–-240, 1934.
\bibitem{af2}
T. G. Wand and C. P. Lee, "Radiation Pressure and Acoustic Levitation," in Nonlinear Acoustics, Chap. 6, edited by M. F. Hamilton and D. T. Blackstock, Academic Press, 1998. 
\bibitem{as1} C. Eckart, "Vortices and Streams Caused by Sound Waves," Physical Review, 73, pp.68--76, 1948.
\bibitem{as2} W. L. Nyborg, "Acoustic Streaming Equations: Laws of Rotational Motion for Fluid Elements," J. Acoust. Soc. Am., 25 (5), pp.938-–944, 1953.
\bibitem{as3} K. Hasegawa, L. Qiu, A. Noda, S. Inoue, and H. Shinoda, "Electronically Steerable Ultrasound-Driven Long Narrow Air Stream," Applied Physics Letters, 111, 064104, 2017.
\bibitem{ps1} P. J. Westervelt, "Parametric Acoustic Array," J. Acoust. Soc. Am., 35 (4), pp. 535-–537, 1963.
\bibitem{ps2} M. Yoneyama, J. Fujimoto, Y. Kawamo, and S. Sasabe, "The Audio Spotlight: An Application of Nonlinear Interaction of Sound Waves to a New Type of Loudspeaker Design," J. Acoust. Soc. Am., 73 (5), pp. 1532-–1536, 1983.
\bibitem{al1}A. Marzo, A. Barnes, and B. W. Drinkwater, "TinyLev: A multi-emitter single-axis acoustic levitator," Rev. Sci. Instrum., 88 (8), 085105, 2017.
\bibitem{al2}I. Ezcurdia, R. Morales, M. A. B. Andrade, and A. Marzo, "LeviPrint: Contactless Fabrication using Full Acoustic Trapping of Elongated Parts," in ACM SIGGRAPH 2022 Conference Proceedings (SIGGRAPH '22), Article 52, pp.1–9, 2022.
\bibitem{al3} S. Inoue, S. Mogami, T. Ichiyama, A. Noda, Y. Makino, and H. Shinoda, "Acoustical boundary hologram for macroscopic rigid-body levitation," Journal of the Acoustical Society of America, 1. 145, pp. 328–337, 2019.
\bibitem{om1} R. Hirayama, G. Christopoulos, D. Martinez Plasencia, and S. Subramanian, 
"High-speed acoustic holography with arbitrary scattering objects," Sci. Adv., 8, Vol 8, Issue 24, 2022.
\bibitem{om2} A. Marzo, S. Seah, B. Drinkwater, et al., "Holographic acoustic elements for manipulation of levitated objects," Nat Commun 6, 8661, 2015.
\bibitem{td1} I. Rakkolainen, E. Freeman, A. Sand, R. Raisamo, and S. Brewster, "A survey of mid-air ultrasound haptics and its applications," IEEE Trans Haptics, vol. 14, no. 1, pp. 2–19, 2021.
\bibitem{td2} D. Hajas, D. Pittera, A. Nasce, O. Georgiou, and M. Obrist, "Mid-air haptic rendering of 2D geometric shapes with a dynamic tactile pointer," IEEE Trans Haptics, vol. 13, no. 4, pp. 806-817, 2020.
\bibitem{td3} R. Takahashi, K. Hasegawa, and H. Shinoda, "Tactile Stimulation by Repetitive Lateral Movement of Midair Ultrasound Focus," IEEE Transactions on Haptics, vol. 13, no. 4, pp. 334-342, 2020.
\bibitem{frag}K. Hasegawa, L. Qiu, and H. Shinoda, "Midair Ultrasound Fragrance Rendering," IEEE Transactions on Visualization and Computer Graphics, vol. 24, no. 4, pp. 1477-1485, 2018.
\bibitem{aerosol}K. Nagata and K. Hasegawa, "Suppression of Short Range Exposure to Infectious Aerosols using Multiple Paths of Midair Ultrasound Acoustic Streaming", Aerosol Science and Technology, pp.1–16. https://doi.org/10.1080/02786826.2024.2345156, 2024.
\bibitem{tatezono}T. Iwamoto, M. Tatezono, and H. Shinoda, "Non-contact Method for Producing Tactile Sensation Using Airborne Ultrasound," in Proc. of Eurohaptics 2008, pp.504-513, 2008.
\bibitem{hoshi}T. Hoshi, M. Takahashi, T. Iwamoto, and H. Shinoda, "Noncontact Tactile Display Based on Radiation Pressure of Airborne Ultrasound," IEEE Trans. on Haptics, Vol. 3, No. 3, pp.155-165, 2010.
\bibitem{aupa1}S. Suzuki, S. Inoue, M. Fujiwara, Y. Makino, and H. Shinoda, "AUTD3: Scalable Airborne Ultrasound Tactile Display," IEEE Transactions on Haptics, Vol. 14, No. 4, pp. 740-749, 31 March 2021.
\bibitem{aupa2}A. Marzo, T. Corkett and B. W. Drinkwater, "Ultraino: An Open Phased-Array System for Narrowband Airborne Ultrasound Transmission," IEEE Transactions on Ultrasonics, Ferroelectrics, and Frequency Control, vol. 65, no. 1, pp. 102-111, 2018.
\bibitem{uneven1}K. Melde, A. G. Mark, T. Qiu, and P. Fischer, "Holograms for Acoustics," Nature, 537, pp.518-522, 2016.
\bibitem{uneven2}S. Polychronopoulos and G. Memoli, "Acoustic Levitation with Optimized Reflective Metamaterials," Scientific Reports, 10, 4254, 2020.
\bibitem{am1} R. Al. Jahdali and Y. Wu, "High Transmission Acoustic Focusing by Impedance-Matched Acoustic Meta-Surfaces," Applied Physics Letters, 108, 031902, 2016.
\bibitem{am2} S. D. Zhao, A. L. Chen, Y. S. Wand, and C. Zhang, "Continuously Tunable Acoustic Metasurface for Transmitted Wavefront Modulation," Phys. Rev. Applied, 10, 054066, 2018.
\bibitem{am3} A.-C. Hladky-Hennion, J. O. Vasseur, G. Haw, C. Cro\"{e}nne, L. Haumesser, and A. N. Norris, "Negative refraction of acoustic waves using a foam-like metallic structure," Appl. Phys. Lett, 102 (14), 144103, 2013.
\bibitem{am4} L. Zhao, E. et al., "Ultrasound beam steering with flattened acoustic metamaterial Luneburg lens," Appl. Phys. Lett., 116 (7), 071902, 2020.
\bibitem{fzp1} D. Schindel, A. Bashford, and D. Hutchins, "Focussing of ultrasonicvwaves in air using a micromachined fresnel zone-plate," Ultrasonics 35, pp.275-285, 1997.
\bibitem{fzp2} X. Xia, Y. Li, F. Cai, H. Zhou, T. Ma, and H. Zheng, "Ultrasonic tunable focusing by a stretchable
phase-reversal Fresnel zone plate," Applied Physics Letters, 117, 021904, 2020.
\bibitem{fzp3} S. P\'{e}rez-Lop\'{e}z, J. M. Fuster, and P. Candelas, "Spatio-temporal
ultrasound beam modulation to sequentially achieve multiple foci with a single planar monofocal lens," Scientic Reports, 11, 13458, 2021.
\bibitem{fzp4} S. Hur, H. Choi, G. H. Yoon, N. W. Kim, D.-G. Lee, and Y. T. Kim, "Planar
ultrasonic transducer based on a metasurface piezoelectric ring array for subwavelength acoustic focusing in water," Scientific Reports 12, 1485, 2022.
\bibitem{kitano}M. Kitano and K. Hasegawa, "Airborne Ultrasound Focusing Aperture with Binary Amplitude Mask Over Planar Ultrasound Emissions," Journal of Applied Physics, 133, 144901, 2023.
\bibitem{plate} G. V. Selicani, and F. Buiochi, "Stepped-plate ultrasonic transducer used as a source of harmonic radiation force optimized by genetic algorithm," Ultrasonics, Vol. 116, 106505, 2021.
\bibitem{plate2}Y. Ito, R. Kato, and A. Osumi, "Examinations of behavior of liquid irradiated with highintensity
aerial ultrasonic waves in a long pore," Jpn. J. Appl. Phys., 52, 7S, 07HE12, 2013.
\end{thebibliography}
\end{document}